\documentclass[twocolumn,prb,showpacs]{revtex4}
\usepackage{graphicx}
\usepackage{dcolumn}
\usepackage{bm}
\usepackage{amsmath}
\usepackage{amssymb}

\begin{document}

\preprint{}
\title{ Supercurrent-induced phonon angular momentum}
\author{Takehito Yokoyama}
\affiliation{Department of Physics, Institute of Science Tokyo, Tokyo 152-8551,
Japan
}
\date{\today}

\begin{abstract}
We propose a mechanism of supercurrent-induced phonon angular momentum in mixed parity superconductors and $s$-wave superconductors with spin orbit coupling. We derive analytical expressions of phonon angular momentum induced by the supercurrent by perturbative calculation. 
The physical interpretation of this effect is also discussed. 
\end{abstract}

\maketitle

\section{Introduction}

Chiral phonons are collective lattice vibrations that possess a definite rotational direction (angular momentum).\cite{LZhang2014,LZhang2015,Zhang2023,Juraschek2025,Zhang2025} Unlike conventional linearly polarized phonons, chiral phonons exhibit circular or elliptical polarization in crystals, arising from the rotational motion of atoms around their equilibrium positions.
This leads to phenomena such as phonon angular momentum, phonon magnetic moments, and chirality-dependent interactions with electrons and photons.\cite{HZhu2018,Ishito2023,Ishito2023b,Grissonnanche2020,XTChen2019,Li2019,Ueda2023,Zhang2024}

The interplay between chiral phonons and electrons has been extensively studied in recent years. Chiral phonons can directly influence spin dynamics through an effective magnetic field induced by chiral phonons.~\cite{Nova2017,Juraschek2019,Geilhufe,Juraschek2022,Xiong2022,Luo2023,Hernandez2022,Chaudhary2023,Merlin2025} Recent studies in this field include conversion between electron spin and microscopic atomic rotation~\cite{HamadaPRR2020}, magnetization manipulation or reversal\cite{Basini,Kahana2023,Davies2023}, exciton formation\cite{Liu2019,Lujan2024}, electron (chirality)-chiral phonon coupling\cite{Tateishi2025,Chen2025},
interactions between chiral phonons and eletronic spin\cite{Fransson2023} or orbital\cite{Ren2021} magnetizations, spin and orbital currents generated by chiral phonons\cite{Kim2023,Li2022,Yao,Funato,Qin2025,Nabei2026}, and  chiral phonon-mediated spin-spin interaction\cite{Korenev,Jeong2022,Yokoyama2024}. 
The coupling between chiral phonons and electron spins opens up new avenues for the design of spintronic devices, where phonon-mediated spin control could complement, or potentially replace, conventional approaches based on external magnetic fields or spin-orbit coupling.

Phonon angular momentum can be induced by external stimuli.
Phonon angular momentum can be induced by the temperature gradient.\cite{Hamada2018} This is reminiscent of the Edelstein effect in electronic systems: current-induced spin angular momentum.\cite{Edelstein1990} 
Also, phonon angular momentum induced by an applied electric field in magnetic insulators has been predicted.\cite{Hamada2020}  This bears a close resemblance to magnetoelectric effects in insulators. 
As for chiral phonons in metals, it has been recently shown that  charge (Ohmic) current can induce phonon angular momentum, which is dubbed phonon Edelstein effect. \cite{Yokoyama2025PhononEdelstein}

In this paper, we investigate a superconducting analog of this effect.
We propose a mechanism of supercurrent-induced phonon angular momentum in mixed parity superconductors and $s$-wave superconductors with spin orbit coupling. We derive analytical expressions of phonon angular momentum induced by the supercurrent by perturbative calculation. 
The physical interpretation of this effect is also discussed.

\begin{figure}[htb]
\begin{center}
\includegraphics[clip,width=5.0cm]{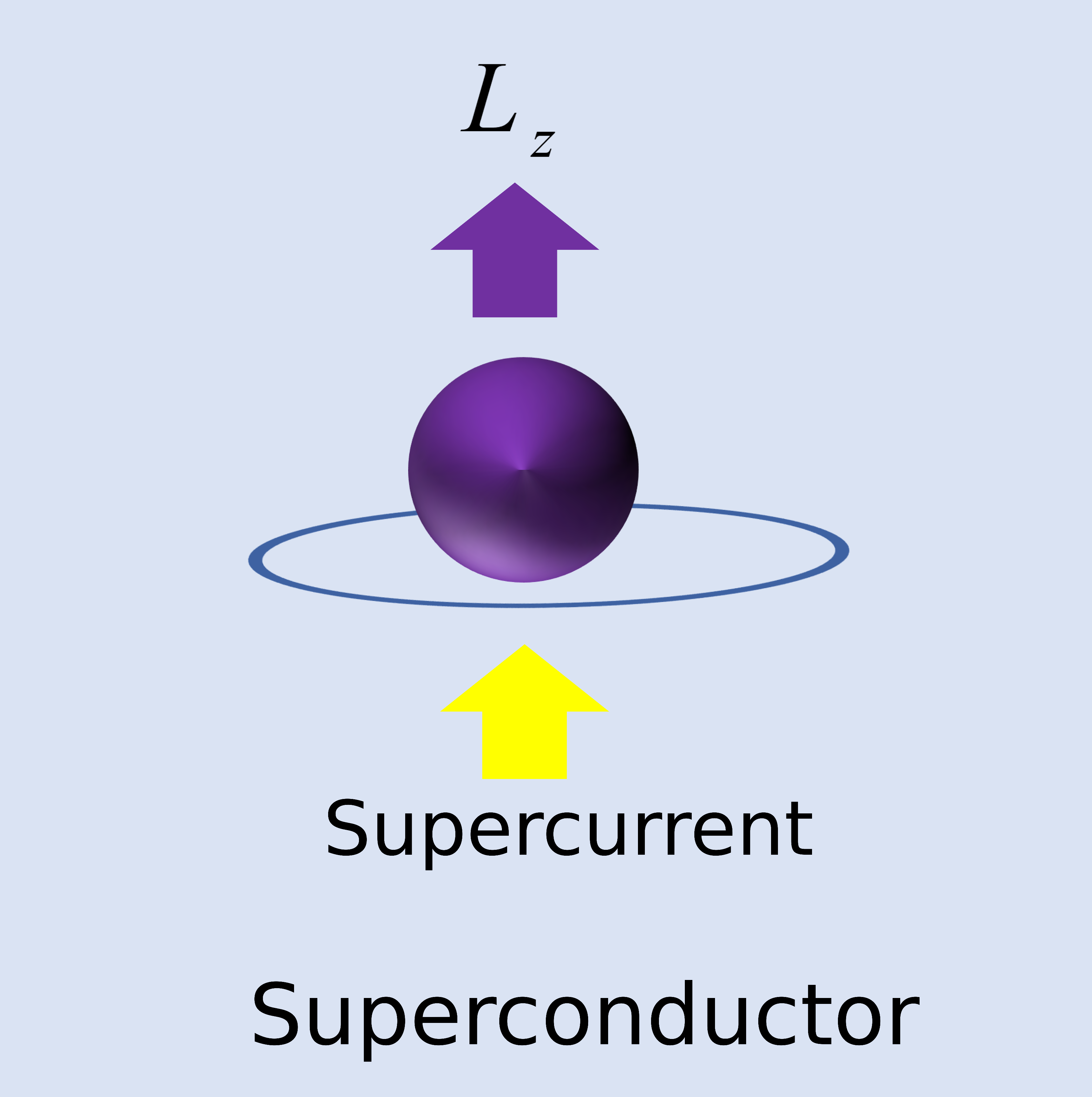}
\end{center}
\caption{
Schematic diagram of the model. The application of supercurrent to a superconductor with chiral phonons can induce phonon angular momentum.}
\label{f1}
\end{figure}

We consider three-dimensional superconductors in the presence of chiral phonons and investigate  the phonon angular momentum in response to a supercurrent as shown in Fig. \ref{f1}.
We assume that the superconductor has a helical crystal structure with the space group $P3_1 21$ or $P3_2 21$ ($D^4_3$ or $D^6_3$) corresponding to the right-handed or left-handed screw symmetry.
 Then, the response tensor $\chi_{ij}$, defined by $L^i=\chi_{ij} A_{j}$, is diagonal: $\chi_{ij} \sim \delta_{ij}$\cite{Hamada2018,Birss1962} where $L^i$ and $A_{j}$ denote phonon angular momentum and a vector potential, respectively. 
In this paper, we study the response of phonon angular momentum along the $z$-axis to supecurrent applied along the $z$-axis.
In general, nonzero $\chi_{ij}$ is allowed for gyrotropic crystals. Among 21 point groups lacking inversion symmetry, 18 point groups are gyrotropic (from which 11 point groups are chiral).\cite{Kizel1975,Jerphagnon1976,Ganichev2019}

\section{Mixed parity superconductors}

We now investigate supercurrent-induced phonon angular momentum in mixed parity superconductors.
We consider the Hamiltonian for phonons $H_p$ as follows
\begin{eqnarray}
{H_p} = \sum\limits_{{\bf{q}}\nu} {\omega _{\bf{q}\nu}} a_{\bf{q}}^{\nu\dag }a_{\bf{q}}^{\nu }
\end{eqnarray}
where $a_{\bf{q}}^{\nu \dag } (a_{\bf{q}}^{\nu})$ is the creation (annihilation) operator for a phonon with wave vector $\bf{q}$ and mode $\nu$, with the corresponding phonon dispersion $\omega _{\bf{q}\nu}$.
We set $k_B=\hbar=1$ throughout this paper. 

We define the phonon Green's function $D$ by
\begin{eqnarray}
D({\bf{q}},t ,\nu) =  - i\left\langle {{{\rm{T}}_\tau }\left[ {{\varphi _{{\bf{q}}\nu}}(t)\varphi _{{\bf{q}}\nu}^\dag } \right]} \right\rangle 
\end{eqnarray}
with $\varphi _{{\bf{q}}\nu}^\dag  = \frac{1}{{ 2 }}\left( {a_{\bf{q}}^\nu + a_{ - {\bf{q}}}^{\nu\dag }, - i\left( {a_{\bf{q}}^{\nu } - a_{ - {\bf{q}}}^{\nu \dag} } \right)} \right)$.
These two components correspond to the second quantization representation of the displacement and momentum of atoms.

We perform a Fourier transform of the phonon Green's function $D({\bf{q}},t ,\nu)$ for $H_p$ and obtain 
\begin{eqnarray}
D = \frac{1}{2(\nu_n^2 + \omega_{q\nu}^2)} \begin{pmatrix} -\omega_{q\nu} & \nu_n \\ -\nu_n & -\omega_{q\nu} \end{pmatrix} \equiv D_0 \rho_0 + D_y i \rho_y
\end{eqnarray}
with  the unit matrix $\rho_0$ and  the Pauli matric $\rho_y$. Here, $\nu_n$ is the bosonic Matsubara frequency.
The angular momentum of phonons along the $z$ direction can be expressed by the second quantization representation of the displacement $u$ and momentum $p$  as \cite{Gao2023}
\begin{eqnarray}
{L^z} = u^xp^y - u^yp^x = \frac{1}{{2}}\sum\limits_{{\bf{q}},\nu} {g_{\bf{q}}^{\nu ,{\nu^\prime }}\left( {a_{\bf{q}}^\nu + a_{ - {\bf{q}}}^{\nu\dag }} \right)} \left( {a_{\bf{q}}^{\nu '\dag } - a_{ - {\bf{q}}}^{\nu'}} \right)
\end{eqnarray}
with
\begin{eqnarray}
g_{\bf{q}}^{\nu ,{\nu^\prime }} = \sqrt {\frac{{\omega _{\bf{q}\nu^\prime }}}{{\omega _{\bf{q}\nu}}}} \xi _{{\bf{q}},{\nu ^\prime }}^\dag \left( {\begin{array}{*{20}{c}}
0&{ - i}\\
i&0
\end{array}} \right){\xi _{{\bf{q}},\nu }}.
\end{eqnarray}
Here, ${\xi _{{\bf{q}},\nu }}$ is the polarization vector. We assume doubly degenerate oppositely circularly polarized phonon modes.\cite{LZhang2015} 
It is straightforward to extend the following results to non degenerate cases by keeping the phonon mode index $\nu$ for the phonon dispersions in the calculations below.
Each of circular polarized phonon modes is labeled by $\nu=\pm$. Namely, we assume
${\xi _{{\bf{q}},\nu }} = \frac{1}{{\sqrt 2 }}{\left( {\begin{array}{*{20}{c}}
1&{\nu  i} \end{array}} \right)^t}$.
Then we have
$g_{\bf{q}}^{\nu ,{\nu ^\prime }} = \nu {\delta _{\nu ,\nu '}}$
and obtain
\begin{eqnarray}
{L^z} = \sum\limits_{{\bf{q}},\nu } {\nu \varphi _{{\bf{q}}\nu }^\dag {\rho_y}\varphi _{{\bf{q}}\nu }^{}}.
\end{eqnarray}

We now turn to the description of the electronic system.
We consider superconductors without inversion symmetry where the parity of the gap function is also broken and treat the following model Hamiltonian for mixed parity superconductors $H_e$ \cite{Yip2014}
\begin{eqnarray}
H_e = \xi \tau_3 \sigma_0 + \tau_1 (\Delta_s \sigma_0 + \mathbf{d} \cdot \boldsymbol{\sigma}) 
\end{eqnarray}
where $\xi=k^2/(2m)-\mu,$ $\Delta_s \in \mathbb{R}$ and $ \mathbf{d} \in \mathbb{R}^3$ are the kinetic energy, the singlet gap function and the $\mathbf{d}$-vector for the triplet gap function, respectively. Here, $\mu$ is the chemical potential. 
Also, $\tau $ and $\sigma$ are Pauli matrices in particle-hole and spin spaces, respectively ($\sigma _0$ is the unit matrix).
Here, we omit the tensor product. For example, ${\tau _3}{\sigma _0}$ should be interpreted as ${\tau _3} \otimes {\sigma _0}$.
Note that we use the basis such that gap functions for singlet pairing are proportional to $\sigma_0$.\cite{Ivanov2006,Yokoyama2025Floquet}
In the following, $\tau_0$ and $\sigma_0$ are omitted in trivial cases.

The Green function is then given by 
\begin{eqnarray}
G=(i \omega_n-H_e)^{-1} \nonumber \\ = \frac{1}{A^2 - |\mathbf{g}|^2} (A +  \mathbf{g} \cdot \boldsymbol{\sigma}) [i \omega_n + \xi \tau_3 + \tau_1 (\Delta_s \sigma_0 + \mathbf{d} \cdot \boldsymbol{\sigma})], 
\end{eqnarray}
with $\mathbf{g} = 2 \Delta_s \mathbf{d}$ and $ A = -\omega_n^2 - \xi^2 - \Delta_s^2 - |\mathbf{d}|^2$. Here, $\omega_n$ is the fermionic Matsubara frequency.

The applied supercurrent along the $z$-direction is also described by 
\begin{eqnarray}
{H_A} =  - {j_z}{A_z}
\end{eqnarray}
where $j_z =-\frac{e}{m} k_z \tau_0 \sigma_0$  and  $A_z$ are, respectively, the current operator  and the vector potential.

We assume the  interaction between  chiral phonons and electrons in the form of spin-orbit coupling:
\begin{eqnarray}
H_{ep} = \lambda L^z \sigma_z, 
\end{eqnarray}
with $L^z = \nu \rho_y$ and $\sigma_z = \tau_0 \sigma_3$. The parameter $\lambda$ represents the coupling strength.
This interaction stems from Zeeman effect by the magnetic field due to chiral phonons. We assume that atoms rotate within the $xy$ plane. Then, the associated magnetic field is parallel to the $z$ axis and hence only $\sigma _z$ is involved in this coupling.
Since this coupling stems from the magnetic field due to rotation of phonons, this coupling would arise for all types of chiral phonons.

\begin{figure}[htb]
\begin{center}
\includegraphics[clip,width=8cm]{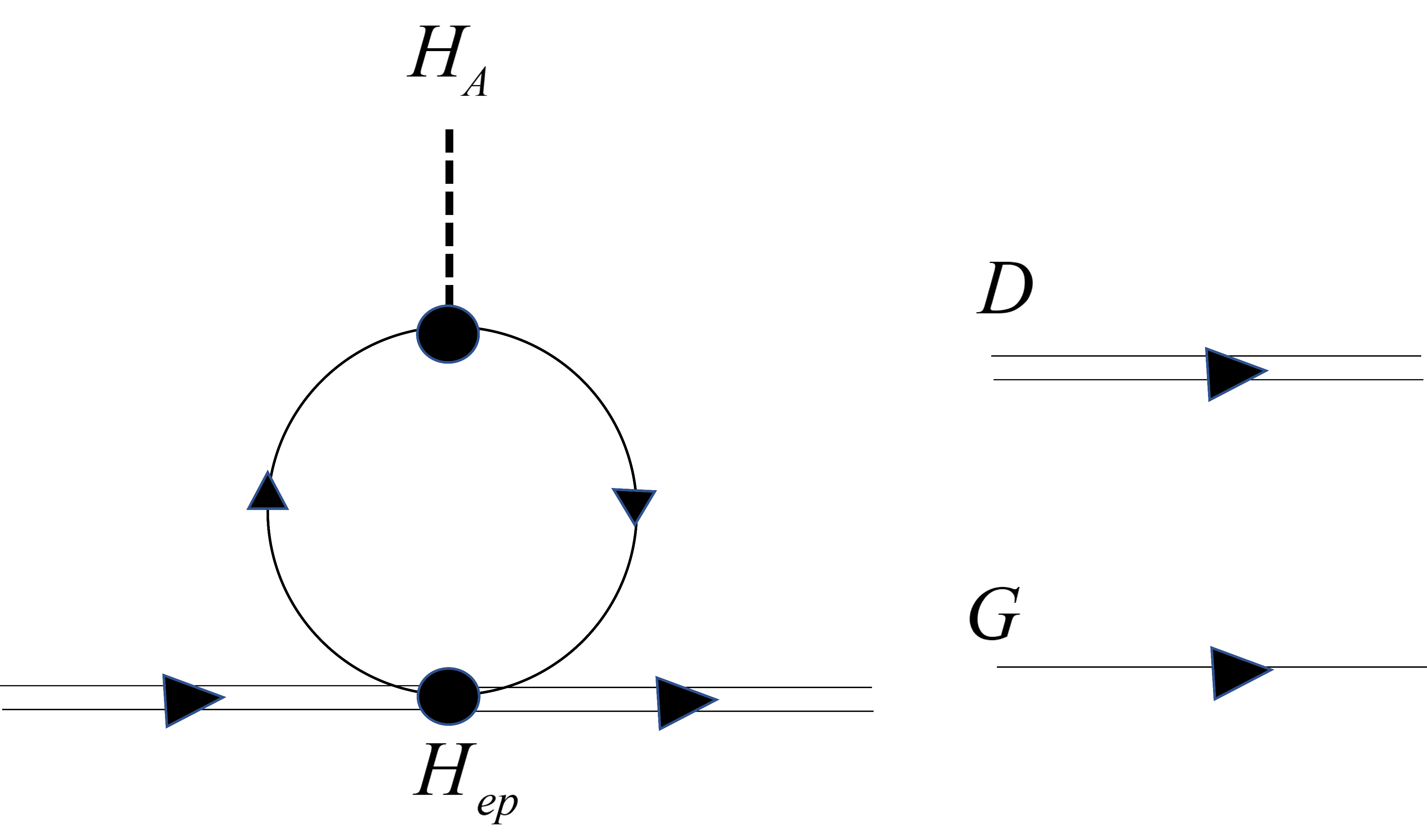}
\end{center}
\caption{
The diagrammatic representations of the phonon Green's function to first orders in ${H_{ep}} $ and  ${H_A} $. Two parallel solid, solid and dotted lines represent phonon Green's function, electron Green's function, and ${H_A} $, respectively. Two parallel solid and solid lines intersect via ${H_{ep}} $.}
\label{f2}
\end{figure}

In the following, we treat ${H_{ep}} $ and  ${H_A} $ as perturbations.
We calculate the expectation value of $L^z$ with respect to the first orders of ${H_{ep}} $ and  $H_A$.
Then, the phonon angular momentum can be calculated as 
\begin{eqnarray}
\langle {L}^z \rangle = -\text{Tr } {L}^z {\cal{D}} = \text{Tr} \left[{L}_z D \lambda L_z \sigma_z G j_z A_z G D \right].
\end{eqnarray}
Here, $\cal{D}$ is the full phonon Green's function. The trace is taken over the phononic and electronic degrees of freedom. The diagrammatic representation of this expansion of the phonon Green's function is shown in Fig. 2.

First, we take trace for the phonon part:
\begin{align}
 \text{Tr}_{\rho,q,n} (\rho_y D \rho_y D) =T \sum_{n,q}  \frac{\omega_q^2 - \nu_n^2}{2(\nu_n^2 + \omega_q^2)^2} \nonumber
\end{align}
\begin{eqnarray}
= \frac{\pi^2}{8 T} \sum_{q} \frac{1}{\sinh^2 \frac{\omega_q}{2T}}
\end{eqnarray}
where $n$ runs over the integers and the subscripts of Tr indicate the degrees of freedom over which the partial trace is taken.

For acoustic phonon with $\omega_q = v_a q$, we can perform the intergration over phonon momemtum: 
\begin{align}
\left( \frac{1}{2\pi} \right)^3 \int_0^\infty \frac{4 \pi q^2 dq}{\sinh^2 \frac{v_a q}{2T}} = \frac{2 T^3}{3 v_a^3}.
\end{align}

Next, we take trace over the electronic degrees of freedom:
\begin{eqnarray}
\text{Tr}_{\sigma,\tau,n,k}\sigma_z G j_z A_z G= - \frac{e}{m} A_z \text{Tr}_{\sigma,\tau,n,k} k_z \sigma_z G^2
\end{eqnarray}
\begin{eqnarray}
= T \sum_{n,k} \frac{4e A_z k_z}{m(A^2 - |\mathbf{g}|^2)^2} \left( -A^2 + \mathbf{g}^2 - 4 A \omega_n^2 \right) \mathbf{g}_z
\end{eqnarray}
where $n \in \mathbb{Z}$.

As an example of the $\mathbf{d}$-vector, 
we now consider $\mathbf{d} = \Delta_t \mathbf{\hat{k}}$ and the tempereture regime near the superconducting transition temperature $T_C$ where $\mathbf{\hat{k}}$ denotes the unit vector along $\mathbf{k}$.
Then, we have 
\begin{eqnarray}
\text{Tr }_{\sigma,\tau,n,k}\sigma_z G j_z A_z G \cong  -\frac{8e A_z }{3m}T \sum_{n, \mathbf{k}} \Delta_s \Delta_t \frac{k(3 \omega_n^2 - \xi^2)}{(\omega_n^2 + \xi^2)^3} \nonumber
\end{eqnarray}
\begin{eqnarray}
 = \frac{4 \pi^5 \Delta_s \Delta_t e A_z}{3m T^2} \int k^3 \frac{\sinh \frac{\pi \xi}{2T}}{\xi \cosh^3 \frac{\pi \xi}{2T}} d k \nonumber
\end{eqnarray}
\begin{eqnarray}
\cong -\frac{4 \pi^5  k_F^2 \Delta_s \Delta_t e A_z}{3 T^2} \int \frac{\sinh \frac{\pi \xi}{2T}}{\xi \cosh^3 \frac{\pi \xi}{2T}} d \xi \nonumber \\ = -\frac{56 \pi^3  k_F^2 e}{3 T^2} \Delta_s \Delta \zeta(3) A_z.
\end{eqnarray}

Therefore, we arrive at the final result of supercurrent-induced phonon angular momentum:
\begin{align}
\langle L^z \rangle =
 - \frac{\pi^2 e \lambda}{2 m} \sum_{q} \frac{1}{\text{sinh}^2 \frac{\omega_{q}}{2T}} \nonumber \\ \times \sum_{k,n} \frac{k_z \mathbf{g}_z}{(A^2 - \mathbf{g}^2)^2} (-A^2 + \mathbf{g}^2 - 4 A \omega_n^2) A_z.
\end{align}

For $\mathbf{d} = \Delta_t \mathbf{\hat{k}}$ and near $T_C$, this can be reduced to 
\begin{align}
\langle L^z \rangle \cong - \frac{7 \pi^5 e \lambda k_F^2 \zeta(3) \Delta_s \Delta_t}{3 T^3} \sum_{q} \frac{1}{\text{sinh}^2 \frac{\omega_{q}}{2T}} A_z.
\end{align}

For acoustic phonon with $\omega_q = v_a q$, we obtain
\begin{align}
\langle L^z \rangle \cong  - \frac{28 \pi^5 e \lambda k_F^2 \zeta(3) \Delta_s \Delta_t}{9 v_a^3} A_z .
\end{align}

Note that the gauge is fixed here and hence the vector potential has a well-defined value. The vector potential can be uniquely determined by solving the Maxwell equations together with the London equation $\bf{j}=\rho \bf{A}$ where $\bf{j}$ and $\rho$ are the supercurrent density  and the superfluid density, respectively.
Gauge invariance in Eqs.(18) and (19) can be restored by replacing $A_z$ with $A_z +\frac{1}{e} \partial_z \phi$ where $\phi$ is  the phase of the superconducting order parameter. 

One can express  Eqs.(18) and (19) as a response to supercurrent using the London equation. By substitution, Eq. (18) can be rewritten as (see Appendix A for details)
\begin{align}
\langle L^z \rangle \cong - \frac{2 \pi^9 m \lambda  \Delta_s \Delta_t}{e k_F T(\Delta_s^2 + \Delta_t^2)} \sum_{q} \frac{1}{\text{sinh}^2 \frac{\omega_{q}}{2T}} j_z
\end{align}
where $j_z$ denotes the supercurrent density along the $z$-direction. 
Similarly, Eq. (19) can be expressed as 
\begin{align}
\langle L^z \rangle \cong  - \frac{8 \pi^9 m \lambda T^2 \Delta_s \Delta_t}{3 v_a^3 e k_F(\Delta_s^2 + \Delta_t^2)} j_z .
\end{align}

Let us estimate the magnitude of Eq. (20). For $\Delta_s \sim \Delta_t$, $T \sim 1$ meV, $\lambda \sim 1$ meV, $k_F/m \sim 10^6$ m/s, $j_z \sim 10^6$ A/cm$^2$, we have 

\begin{align}
 \frac{2 \pi^9 m \lambda  \Delta_s \Delta_t}{e k_F T(\Delta_s^2 + \Delta_t^2)} j_z \sim 2 \times 10^{-3} \hbar \; \AA^{-3}.
\end{align}
Here, we have recoverd $\hbar$.
For $\omega_{q} \sim T$ and the area of the Brillouin zone $\sim \pi^2$, we have $\sum_{q} \frac{1}{\text{sinh}^2 \frac{\omega_{q}}{2T}} \sim 40$.

\section{$s$-wave superconductors with spin orbit coupling}

Here, we consider supercurrent-induced phonon angular momentum in $s$-wave superconductors with spin orbit coupling. The Hamiltonian is composed of two parts:
\begin{align}
H_0 = \xi \tau_3 + \Delta \tau_1, \quad H_\alpha = \alpha \tau_3 (\mathbf{k} \cdot \boldsymbol{\sigma}).
\end{align}
 The $\Delta$ and $\alpha $ represent the $s$-wave gap function and the spin orbit coupling strength, respectively.
We apply supercurrent along the $z$-axis. Then, the vector potential is introduced through the minimal substitution:
\begin{align}
H_A^1 = \frac{e}{m} k_z A_z, \quad H_A^2 = e \alpha A_z \sigma_z.
\end{align}
The $H_\alpha$ represents Weyl type spin-orbit coupling and breaks inversion and mirror symmetries.\cite{Burkov2018,Armitage2018}

Then, the unperturbed Green function $\mathcal{G}$ reads
\begin{align}
\mathcal{G} = (i\omega_n - H_0)^{-1} = -\frac{i\omega_n + \xi \tau_3 + \Delta \tau_1}{\omega_n^2 + \xi^2 + \Delta^2}.
\end{align}

As for the phonon part, we consider the same Hamiltonian as in the previous section. 
We now treat the spin orbit coupling strength $\alpha $ and the vector potential $A_z$ as perturbations. Below, we focus on the contribution from $A_z$  at the first order.

\subsection{First order in $\alpha$}

\begin{figure}[htb]
\begin{center}
\includegraphics[clip,width=9cm]{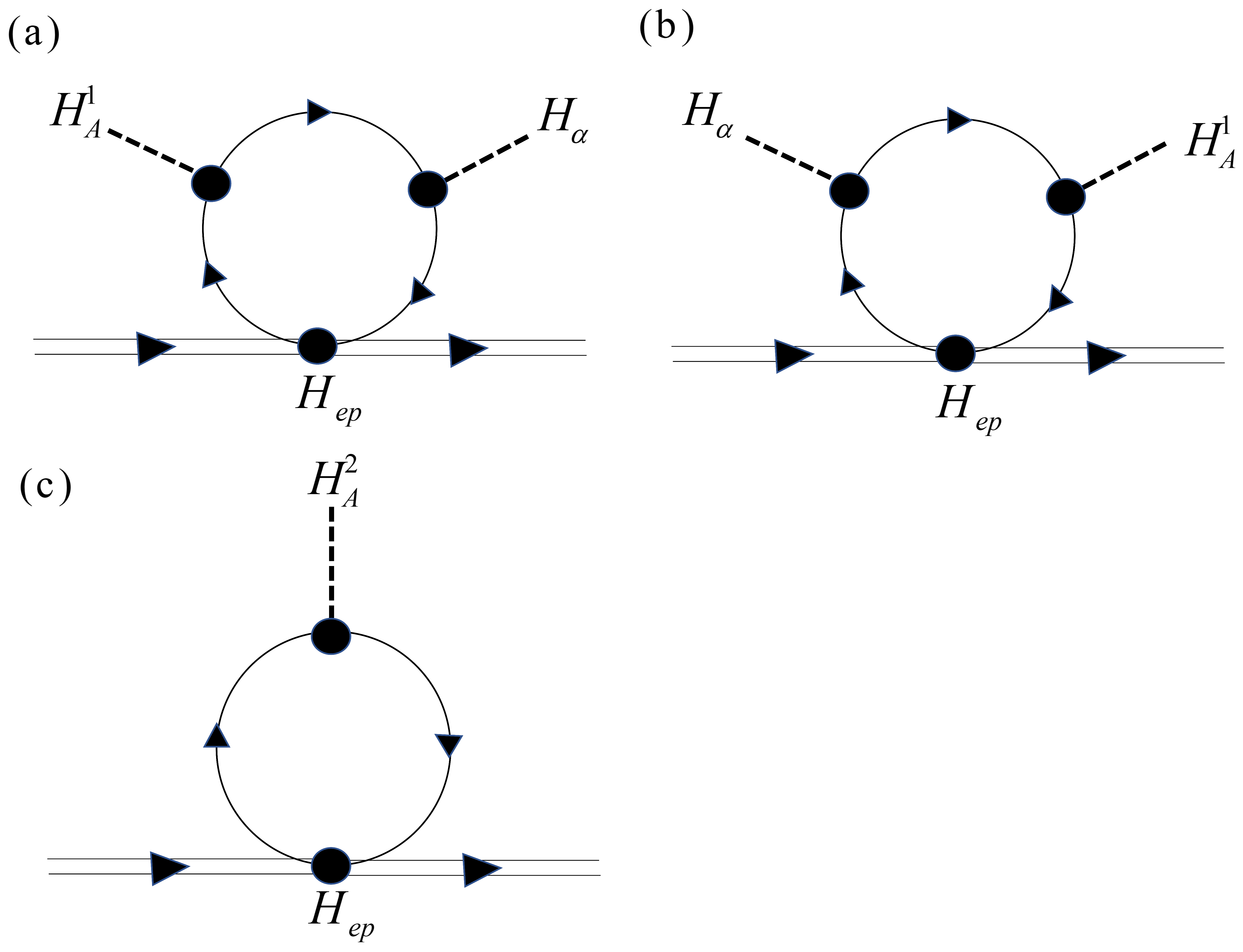}
\end{center}
\caption{
The diagrammatic representations of the phonon Green's function to first orders in $\alpha$, $A_z$, and ${H_{ep}}$. Two parallel and solid lines represent phonon and electron Green's functions, respectively.  Two parallel solid and solid lines intersect via ${H_{ep}}$.  }
\label{f3}
\end{figure}

Let us first consider the terms to the first order in $\alpha$. The corresponding diagrammatic representations are shown in Fig. 3. Since the phonon part is the same as the previous section, we here focus on the electron part. The relevant trace can be calculated as 
\begin{widetext}
\begin{align}
\text{Tr}_{\sigma,\tau} \sigma_z ( \mathcal{G} H_A^1 \mathcal{G} H_\alpha \mathcal{G} + \mathcal{G} H_\alpha \mathcal{G} H_A^1 \mathcal{G} )+ \text{Tr}_{\sigma,\tau} \sigma_z \mathcal{G} H_A^2 \mathcal{G} \nonumber 
\end{align}
\begin{align}
= \alpha A_z \text{Tr }_{\sigma,\tau} \frac{e}{m} k_z^2 \sigma_z ( \mathcal{G}^2 \tau_z \sigma_z \mathcal{G} + \mathcal{G} \tau_z \sigma_z \mathcal{G}^2 )
+ \alpha e A_z \text{Tr }_{\sigma,\tau} \sigma_z  \mathcal{G} \sigma_z \mathcal{G}\nonumber  \\
= 2 e \alpha A_z \text{Tr}_{\tau} \frac{k_z^2}{m} (\mathcal{G}^2 \tau_z \mathcal{G} + \mathcal{G} \tau_z \mathcal{G}^2 )
+ 2 e \alpha A_z \text{Tr}_{\tau} \mathcal{G}^2 \nonumber \\
= 2 e \alpha A_z \text{Tr}_{\tau} \frac{\partial}{\partial k_z} (k_z \mathcal{G}^2).
\end{align}
\end{widetext}
Note that $\frac{\partial \mathcal{G}}{\partial k_z} = \frac{k_z}{m} \mathcal{G} \tau_z \mathcal{G}$.
Thus, this term vanishes by the integration over $k_z$.

\subsection{Third order in $\alpha$}

\begin{figure}[htb]
\begin{center}
\includegraphics[clip,width=6.5cm]{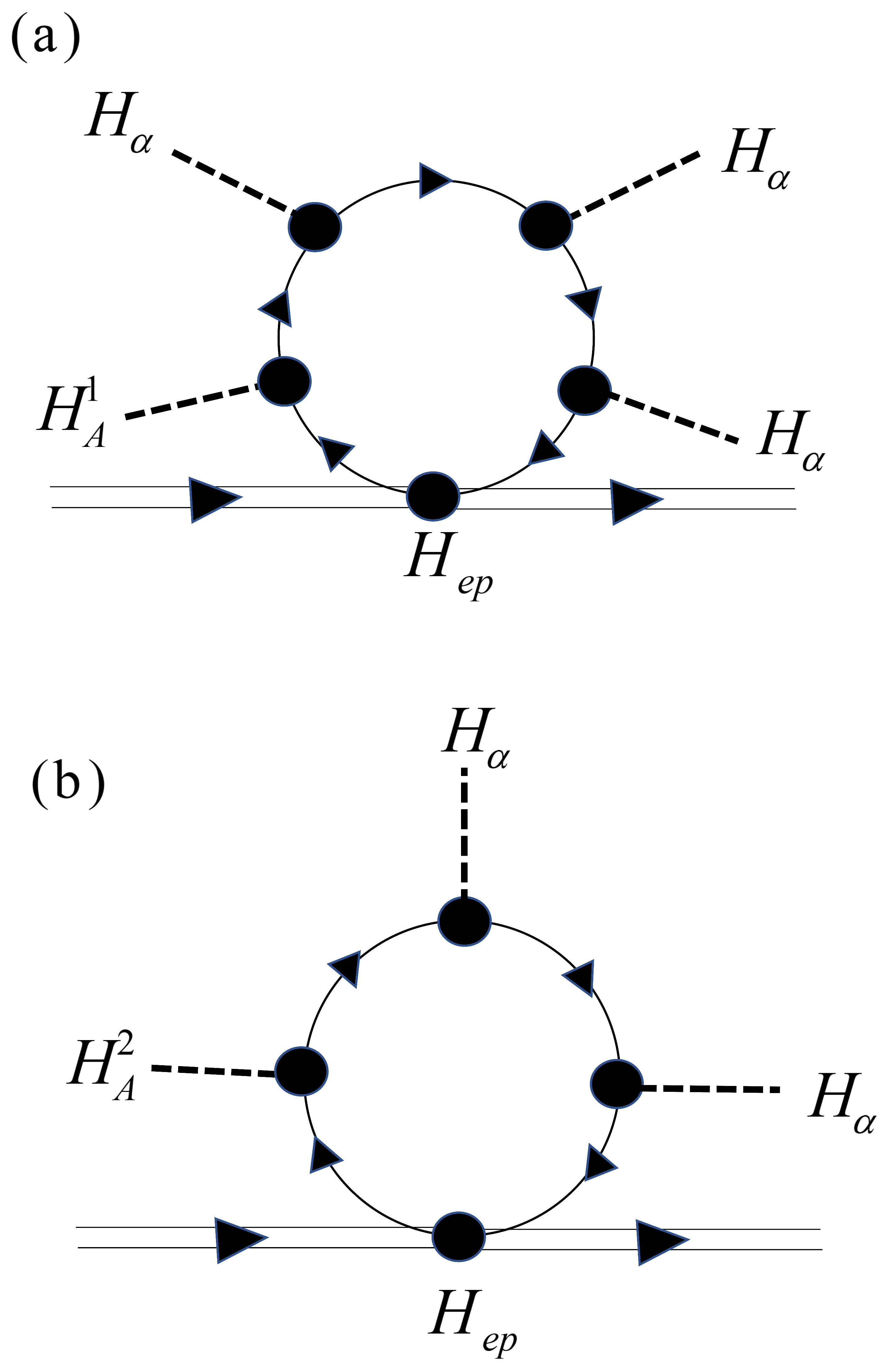}
\end{center}
\caption{
The diagrammatic representations of the phonon Green's function to third order in $\alpha$ and first orders in ${H_{ep}} $ and  $A_z$. Two parallel and solid lines represent phonon and electron Green's functions, respectively.  Two parallel solid and solid lines intersect via ${H_{ep}}$.  }
\label{f4}
\end{figure}

Let us consider the terms of third order in $\alpha$ (the second-order terms in $\alpha$ vanish). The diagrammatic representations are shown in Fig. 4. There are two types of the diagrams: (a) first orders in ${H_A^1}$ and  the third order in ${H_{\alpha}}$  and (b) first orders in ${H_A^2}$ and  the second order in ${H_{\alpha}}$. In the calculatons, cyclic permutations of these purturbations should be also included.
Therefore, the relevant trace for electrons in Fig. 4(a) can be calculated as 
\begin{widetext}
\begin{align}
\text{Tr}_{\sigma,\tau,n,k} \sigma_z ( \mathcal{G} H_A^1 \mathcal{G} H_\alpha \mathcal{G} H_\alpha \mathcal{G} H_\alpha \mathcal{G} + \mathcal{G} H_\alpha \mathcal{G} H_A^1 \mathcal{G} H_\alpha \mathcal{G} H_\alpha \mathcal{G} + \mathcal{G} H_\alpha \mathcal{G} H_\alpha \mathcal{G} H_A^1  \mathcal{G} H_\alpha \mathcal{G} + \mathcal{G} H_\alpha \mathcal{G} H_\alpha \mathcal{G} H_\alpha \mathcal{G} H_A^1 \mathcal{G} ) \nonumber
\end{align}
\begin{align}
= \frac{e \alpha^3}{m} A_z \text{Tr}_{\sigma,\tau,n,k} k_z \sigma_z \left( \mathcal{G}^2 (\mathbf{k} \cdot \boldsymbol{\sigma}) \tau_z \mathcal{G} (\mathbf{k} \cdot \boldsymbol{\sigma}) \tau_z \mathcal{G} (\mathbf{k} \cdot \boldsymbol{\sigma}) \tau_z \mathcal{G} + \text{cyclic permutations} \right) \nonumber
\end{align}
\begin{align}
=\frac{2 e^3 \alpha^3}{m}\, A_z\,
\mathrm{Tr}_{\tau,n,k}\,
k_z^{2} k^{2}
\Big(
 \mathcal{G}^{2}\tau_z \mathcal{G} \tau_z \mathcal{G} \tau_z \mathcal{G}  + \mathcal{G} \tau_z \mathcal{G}^2 \tau_z \mathcal{G} \tau_z \mathcal{G} + \mathcal{G} \tau_z \mathcal{G} \tau_z \mathcal{G}^{2} \tau_z \mathcal{G}  + \mathcal{G} \tau_z \mathcal{G} \tau_z \mathcal{G} \tau_z \mathcal{G}^{2}
\Big)  \nonumber
\end{align}
\begin{align}
= \frac{16 e \alpha^3}{m} A_z T \sum_{k,n} k_z^2 \mathbf{k}^2 \frac{\xi (\Delta^4-4\Delta^2 \omega_n^2 - 5 \omega_n^4 + 10 \omega_n^2 \xi^2 - \xi^4)}{(\omega_n^2 + \xi^2 + \Delta^2)^5}.
\end{align}

Near $T_C$, we focus on terms stemming from Cooper pairs (which vanish for $\Delta=0$). Then, this is approximated as 
\begin{align}
 - \frac{64 e \alpha^3}{m} A_z T \sum_{k,n} k_z^2 \mathbf{k}^2 \frac{\xi \Delta^2 \omega_n^2}{(\omega_n^2 + \xi^2)^5}.
\end{align}

Next, let us consider terms with $H_A^2$ as described in Fig. 4(b).
The relevant trace for electrons  can be calculated as 
\begin{align}
\text{Tr }_{\sigma,\tau,n,k} \sigma_z (\mathcal{G} H_A^2 \mathcal{G} H_\alpha \mathcal{G} H_\alpha \mathcal{G} + \text{cyclic permutations})\nonumber
\end{align}
\begin{align}
= e \alpha^3 A_z \text{Tr }_{\sigma,\tau,n,k} \sigma_z (\mathcal{G} \sigma_z \mathcal{G} (\mathbf{k} \cdot \boldsymbol{\sigma}) \tau_3 \mathcal{G} (\mathbf{k} \cdot \boldsymbol{\sigma}) \tau_3 \mathcal{G} + \text{cyclic permutations})\nonumber
\end{align}
\begin{align}
= 2 e \alpha^3 A_z \text{Tr}_{\tau,n,k} k_z^2 \left( \mathcal{G}^2 \tau_3 \mathcal{G} \tau_3 \mathcal{G} + \mathcal{G} \tau_3 \mathcal{G}^2 \tau_3 \mathcal{G} + \mathcal{G} \tau_3 \mathcal{G} \tau_3 \mathcal{G}^2 \right)\nonumber
\end{align}
\begin{align}
= 4 e \alpha^3 A_z T\sum_{\mathbf{k},n} k_z^2 \frac{(-2\Delta^4 + 4 \Delta^2 (\omega_n^2 + \xi^2) + 6(\omega_n^4 - 6\omega_n^2 \xi^2 + \xi^4))}{(\omega_n^2 + \xi^2 + \Delta^2)^4}.
\end{align}
\end{widetext}

Near $T_C$, we again focus on terms stemming from Cooper pairs (which vanish for $\Delta=0$). Then, for a fixed $n$, this term can be approximated as 
\begin{align}
 16 e \alpha^3 A_z \sum_{\mathbf{k}} k_z^2 \frac{\Delta^2}{(\omega_n^2 + \xi^2)^3}.
\end{align}
The integration over $k$ can be performed as 
\begin{align}
\sum_k k_z^2 k^2 \frac{\xi}{(\omega_n^2 + \xi^2)^5} = \frac{1}{3} \frac{4\pi}{(2\pi)^3} \int \frac{\xi k^6 dk}{(\omega_n^2 + \xi^2)^5} \nonumber\\
\approx \frac{k_F}{6\pi^2} \int 4m^3 \frac{\xi(\xi+\mu)^2}{(\omega_n^2 + \xi^2)^5} d\xi \nonumber\\
= \frac{m^3 \mu k_F}{3\pi^2} \cdot \frac{5\pi}{128 |\omega_n|^7} = \frac{m^3 \mu k_F}{384 \pi} \frac{1}{|\omega_n|^7}
\end{align}
with $k^2= 2m(\xi+\mu)$, and
\begin{align}
\sum_k k_z^2 \frac{1}{(\omega_n^2 + \xi^2)^3} = \frac{1}{6\pi^2} \int \frac{k^4 dk}{(\omega_n^2 + \xi^2)^3} \nonumber\\
\approx \frac{k_F}{6\pi^2} \int \frac{2m^2(\xi+\mu)}{(\omega_n^2 + \xi^2)^3} d\xi \nonumber \\
= \frac{m^2 \mu k_F}{8\pi} \frac{1}{|\omega_n|^5}.
\end{align}

By adding Eqs.(25) and (27) and summing over $n$, we obtain
\begin{align}
\frac{341 \zeta(5)}{96 \pi^6 T^5} e \alpha^3 \Delta^2 m^2 \mu k_F A_z.
\end{align}

Therefore, we arrive at the final expression near $T_C$
\begin{eqnarray}
\langle L^z \rangle = \frac{341 \zeta(5)}{1152 \pi^7 T^5} e \alpha^3 \lambda \Delta^2 m^2 k_F \mu \sum_{q} \frac{1}{\sinh^2 \frac{\omega_{q}}{2T}} A_z
\end{eqnarray}
indicating that supercurrent can induce phonon angular momentum in $s$-wave superconductors with spin orbit coupling.

As for acoustic phonons with $\omega = v_a q$, we obtain
\begin{align}
\langle L^z \rangle = \frac{341 \zeta(5)}{1152 \pi^4 T^2 v_a^3} e \alpha^3 \lambda \Delta^2 m^2 k_F \mu A_z.
\end{align}

Let us express  Eqs. (34) and (35) as a response to supercurrent by means of the London equation.  Eq. (34) can be rewritten as (see Appendix A for details)
\begin{eqnarray}
\langle L^z \rangle = \frac{341 \zeta(5)}{1344 \zeta(3)\pi^3 e T^3} \alpha^3 \lambda m^2 \sum_{q} \frac{1}{\sinh^2 \frac{\omega_{q}}{2T}} j_z.
\end{eqnarray}

Similary, Eq. (35) can be rewritten as 
\begin{align}
\langle L^z \rangle = \frac{341 \zeta(5)}{1344 \zeta(3) v_a^3 e} \alpha^3 \lambda  m^2 j_z.
\end{align}

Let us estimate the magnitude of Eq. (36). For $\alpha \sim 1$ eV $\AA$, $T \sim 1$ meV, $\lambda \sim 1$ meV, $k_F \sim 1 \AA^{-1}$, $k_F/m \sim 10^6$ m/s, $j_z \sim 10^6$ A/cm$^2$, we have 

\begin{eqnarray}
 \frac{341 \zeta(5)}{1344 \zeta(3)\pi^3 e T^3} \alpha^3 \lambda m^2 j_z \sim 9 \times 10^{-6} \hbar \; \AA^{-3}.
\end{eqnarray}
Here, we have recoverd $\hbar$.

The physical interpretation of this effect is as follows.
According to the Edelstein effect, the spin polarization is induced by an application of supercurrent in $s$-wave superconductors with spin orbit coupling and mixed parity superconductors, i.e., $\left\langle {{\sigma _z}} \right\rangle \sim A$.\cite{Edelstein1995,HeLaw2020,Yokoyama2025Nonunitary}
By the mean field approximation, the electron-chiral phonon coupling becomes 
\begin{eqnarray}
{H_{ep}} = \lambda {L^z}\left\langle {{\sigma _z}} \right\rangle.
\end{eqnarray}
Therefore, the phonon angular momentum stems from an orbital Zeemann effect with an effective magnetic field $\left\langle {{\sigma _z}} \right\rangle$ which is proportional to the supercurrent (by assuming the London equation $\bf{j}=\rho \bf{A}$).

Regarding experimental signatures of this effect, we propose that by examining the difference in circularly polarized Raman scattering with and without supercurrent, one can detect the supercurrent-induced phonon angular momentum.

We have clarified the mechanism of phonon angular momentum induced by a vector potential (or phase gradient) in bulk superconductors. However, this mechanism also works in Josephson junctions under the phase bias. Recently, chiral phonons have been observed in a chiral crystal tellurium.\cite{Ishito2023b} Therefore, Nb/Te/Nb Josephson junction is a candidate setup to verify our prediction.

\section{summary}
In this paper, 
we have proposed a mechanism of supercurrent-induced phonon angular momentum in mixed parity superconductors and $s$-wave superconductors with spin orbit coupling. We have derived analytical expressions of phonon angular momentum induced by the supercurrent by perturbative calculation. 
The physical interpretation of this effect is also discussed. 

This work was supported by JSPS KAKENHI Grant No.~JP30578216.

\appendix 
\section{Calculation of superfluid density}

Here, we calculate superfluid densities for $s$-wave and  mixed parity superconductors.
The superfluid density $\rho$  appears as a response of supercurrent to the vector potential:  $j_z = \rho A_z$.
The superfluid density $\rho$ for $s$-wave superconductors can be calculated as\cite{Abrikosov1963}
\begin{align}
    \rho &= \frac{e^2}{m^2} T \sum_{k,n} \text{Tr}_{\tau} \, k_z \mathcal{G} k_z \mathcal{G} - \frac{e^2}{m} N \\
    &= \frac{2 e^2 T}{m^2} \sum_{k,n} k_z^2 \frac{\Delta^2}{(\omega_n^2 + \xi^2 + \Delta^2)^2}
\end{align}
where $\mathcal{G}$ is given in Eq.(25) and $N$ is the number of electrons. Note  that the anomalous Green's function is givwn by $f =- \frac{\Delta}{\omega_n^2 + \xi^2 + \Delta^2} \tau_1$.

Performing the momentum integration and focusing on the temperature near $T_c$, we obtain
\begin{align}
    \rho &= \frac{\pi e^2 k_F^2 T N_F}{3 m^2} \sum_n \frac{\Delta^2}{(\omega_n^2 + \Delta_s^2)^{3/2}} \\
    &\cong \frac{\pi e^2 k_F^2 T N_F}{3 m^2} \sum_n \frac{\Delta^2}{|\omega_n|^3} \\
    &= \frac{7 e^2 k_F^2 N_F \Delta^2 \zeta(3)}{12 m^2 \pi^2 T^2}
\end{align}
with the density of states at the Fermi level $N_F = \frac{m k_F}{\pi^2}$. Thus, we arrive at the expression of the superfluid density $\rho$:
\begin{equation}
    \rho = \frac{7 e^2 k_F^3 \zeta(3) \Delta^2}{12 \pi^4 m T^2}.
\end{equation}

Next, we consider mixed parity superconductors near $T_C$. The calculation is almost the same as that for $s$-wave superconductors. 
Noting that the anomalous Green's function near $T_C$ is given by
\begin{equation}
f \cong - \frac{1}{\omega_n^2 + \xi^2} \tau_1 (\Delta_s \sigma_0 + \bm{d} \cdot \bm{\sigma})
\end{equation}
and the trace over particle-hole and spin spaces gives
\begin{equation}
    \text{Tr}_{\tau,\sigma} f^2 = \frac{4}{(\omega_n^2 + \xi^2)^2} (\Delta_s^2 + |\bm{d}|^2)
\end{equation}

we obtain the final expression for the superfluid density  near $T_C$:
\begin{equation}
 \rho = \frac{7 e^2 k_F^3 \zeta(3)}{6 \pi^4 m T^2} (\Delta_s^2 + |\bm{d}|^2).
\end{equation}

\end{document}